\documentclass[twocolumn,superscriptaddress,amsmath,amssymb,aps,prl,floatfix]{revtex4-2}

\usepackage{graphicx}
\usepackage{bm}

\usepackage[utf8]{inputenc}
\usepackage[T1]{fontenc}

\usepackage{braket}
\usepackage{xcolor}

\usepackage[normalem]{ulem}

\def\Rb87{^{87}\mathrm{Rb}}                             
\def\p0{p_{\mathrm 0}}                             
\def\spinup{\ket{\uparrow}}                             
\def\spindown{\ket{\downarrow}}                        

\def\ex{\mathbf{e}_x}  
\def\ey{\mathbf{e}_y}  
\def\ez{\mathbf{e}_z}  

\begin{document}

\title{Imaginary gauge potentials in a non-Hermitian spin-orbit coupled quantum gas}

\author{J.~Tao}
\affiliation{Joint Quantum Institute, University of Maryland and National Institute of Standards and Technology, College Park, Maryland 20742, USA}
\affiliation{Department of Physics, Harvard University, Cambridge, Massachusetts 02138, USA}
\author{E.~D.~Mercado-Gutierrez}
\affiliation{Joint Quantum Institute, University of Maryland and National Institute of Standards and Technology, College Park, Maryland 20742, USA}
\author{M.~Zhao}
\affiliation{Joint Quantum Institute, University of Maryland and National Institute of Standards and Technology, College Park, Maryland 20742, USA}
\author{I.~B.~Spielman}
\affiliation{Joint Quantum Institute, University of Maryland and National Institute of Standards and Technology, College Park, Maryland 20742, USA}
\email{ian.spielman@nist.gov}

\date{\today} 

\begin{abstract}
In 1996, Hatano and Nelson proposed a non-Hermitian lattice model containing an imaginary Peierls phase [Phys.~Rev.~Lett.~{\bf 77}~570--573 (1996)], which subsequent analyses revealed to be an instance of a new class of topological systems. 
Here, we experimentally realize a continuum analog to this model containing an imaginary gauge potential using a homogeneous spin-orbit coupled Bose-Einstein condensate (BEC).
Non-Hermiticity is introduced by adding tunable spin-dependent loss via microwave coupling to a subspace with spontaneous emission.
We demonstrate that the resulting Heisenberg equations of motion for position and momentum depend explicitly on the system's phase-space distribution.
First, we observe collective nonreciprocal transport in real space, with a ``self-acceleration'' that decreases with the BEC’s spatial extent, consistent with non-Hermitian Gross-Pitaevskii simulations.
We then examine localized edge states: the relatively strong interactions in our BEC suppress the formation of topological edge states, yielding instead highly excited states localized by an interplay between self-acceleration and wavefunction spreading. 
Finally, we confirm that our non-Hermitian description remains valid at all times by comparing to a multi-level master-equation treatment.
\end{abstract}

\maketitle

Open quantum systems are characterized by an interplay between their closed system dynamics and those resulting from the external reservoir.
Because this evolution goes beyond what is possible in closed systems, open systems offer new opportunities and paradigms for quantum engineering~\cite{happer1972optical,briegel2009measurement,terhal2015quantum}.
In some limits, this evolution can be described by effective non-Hermitian Hamiltonians with unusual spectral features such as singular exceptional points~\cite{miri2019exceptional,ashida2020non,bergholtz2021exceptional,ren2022chiral,liu2024third} and phases characterized by spectral topology~\cite{ashida2020non,kawabata2019symmetry,wang2021topological}.
The Hatono-Nelson (HN) model~\cite{hatano1996localization,hatano1997vortex} describes a non-Hermitian lattice with a constant imaginary Peierls phase with topology quantified by an integer winding number~\cite{ashida2020non,kawabata2019symmetry,zhang2020correspondence}.
The resulting non-Hermitian skin effect (NHSE)~\cite{yao2018edge,song2019non2} has been observed for non-interacting particles in a range of experimental platforms~\cite{liang2022dynamic,zhang2021observation,xiao2020non,weidemann2020topological,zhao2025two}.
By contrast with a uniform real-valued Peierls phase, which can be eliminated by a gauge transformation, its imaginary counterpart has direct physical consequences such as non-reciprocal wavefunction localization~\cite{hatano1996localization, yao2018edge,helbig2020generalized,xiao2020non,weidemann2020topological,zhang2021observation} and an equivalent description of evolution on curved space~\cite{Lv2022,Lv2024}.

\begin{figure}[tbh]
\centering
\includegraphics{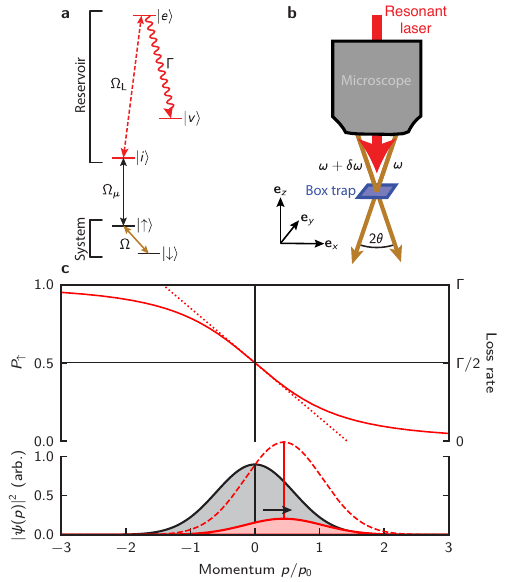}
\caption{Concept.
{\bf a} Level diagram for creating non-Hermitian SOC in a system subspace linked to a reservoir subspace with a microwave field of strength $\Omega_{\mu}$.
Dissipation is induced when the excited state $\ket{e}$ decays with rate $\Gamma$.
{\bf b} Geometry. A homogeneous BEC was confined in a box-shaped trap in the $\ex$-$\ey$ plane, and Raman-dressed by a pair of laser beams projected through a high-resolution microscope objective.
Dissipation is induced by a resonant laser.
{\bf c} Imaginary gauge potential. Left: Spin up population and imaginary gauge field as a function of momentum (solid) along with their linear approximations (dashed).
Right: corresponding momentum dependent loss.  Both computed for $\Omega / \omega_0 = 5.7$.
Bottom: Gaussian momentum distribution before (black) and accelerated after (red) evolving with $\gamma = 0.19 \omega_0$ for $10\ {\rm ms}$.
The solid red curve plots the final distribution including the overall loss term in Eq.~\eqref{eq:imaginary_gauge_potential}, while the dashed curve includes only the imaginary gauge potential.
}
\label{fig:setup}
\end{figure}

We experimentally studied an interacting atomic Bose-Einstein condensate (BEC) the continuum analog subject to the (disorder free) HN model in real space.
This model features a spatially uniform imaginary gauge potential with single particle non-Hermitian Hamiltonian
\begin{align}
    \hat{H}_{\text{nh}} &= \frac{(\hat{p} - i {\mathcal B})^2}{2m^*} + V(\hat x) -i \hbar \frac{\gamma}{2} \label{eq:imaginary_gauge_potential},
\end{align}
with effective mass  $m^*$, canonical momentum $\hat p$, imaginary gauge potential ${\mathcal B}$, external potential $V$, and since there is no gain in our atomic BEC, an overall loss rate $\gamma$.
We introduced $\mathcal B$ into atomic BEC's by combining laser-induced spin-orbit coupling (SOC) and spin-dependent loss.
As shown in Fig.~\ref{fig:setup}{\bf a}-{\bf b} a pair of lasers Raman-coupled two internal states $\ket{\downarrow}$ and $\ket{\uparrow}$ with strength $\Omega$, recoil momentum $\p0$~\cite{lin2011spin} and recoil energy $\hbar \omega_0 = p^2_0/ (2m)$; an additional microwave field drove transitions between $\ket{\uparrow}$ and a lossy ``reservoir'' subspace. 
This gives rise to a two-band non-Hermitian SOC model with Hamiltonian
\begin{equation}
\label{eq:non-Hermitian_Hamiltonian}
    \hat H_{\rm SOC} = \frac{ (\hat{p} + \p0 \sigma_z)^2 }{2m} + \frac{\hbar \Omega}{2}\sigma_x +\frac{\hbar\delta}{2} \sigma_z - i \hbar \gamma \mathcal{P}_{\uparrow}
\end{equation}
in terms of Pauli operators $\sigma_{x,y,z}$, quasi-momentum $\hat{p}$~\cite{lin2009bose}, atomic mass $m$, detuning $\delta$, and spin projection operator $\mathcal{P}_{\uparrow} = \ket{\uparrow}\!\bra{\uparrow}$.

We make an adiabatic approximation in which dynamics are projected into the lowest band and focus on the large Raman coupling ($\Omega > 4 \omega_0$) ``single minimum'' limit, 
where the Hermitian contribution to $\hat H_{\rm SOC}$ has low-energy dynamics well described by the effective mass $m^*$~\cite{Lin2011a}.
The ground state, a plane-wave at $p=0$, consists of an equal superposition of $\spinup$ and $\spindown$; for eigenstates with $p \ll \sqrt{2 m \hbar\Omega}$ the occupation probabilities are proportional to $p$ with $P_{\uparrow,\downarrow}(p) = [1\mp 2 p_0 p / (m \hbar\Omega)]/2$.
As can be seen in Fig.~\ref{fig:setup}{\bf c}, these probabilities lead to a lowest band non-Hermitian contribution to Eq.~\eqref{eq:non-Hermitian_Hamiltonian} of $-i \hbar \gamma P_{\uparrow}(\hat{p})$, equivalent to an imaginary gauge field
\begin{align}
\mathcal{B} &= - p_0 \frac{\gamma}{\Omega-4\omega_0}\label{eq:gauge}
\end{align}
along with an overall loss rate $\gamma/2$ (in what follows we will omit this straightforward contribution).
These expressions are suitable for our experimental configuration (see Methods).

{\it Setup} -- We realized this configuration for nearly pure $\Rb87$ BECs (with $N_0\approx 2\times 10^5$ atoms and chemical potential $\mu \approx h\times 500\ {\rm Hz}$) confined by the joint action of blue and red detuned optical dipole traps (ODTs, with wavelengths $635\ {\rm nm}$ and $1064\ {\rm nm}$ respectively).
The blue ODT horizontally confined the BEC in a $L=30\ \mu{\rm m}$ square (in the $\ex$-$\ey$ plane, projected using a numerical aperture ${\rm NA}=0.55$ microscope) while the red ODT provided a vertically oriented ($\ez$) harmonic potential with frequency $\omega_z /(2\pi) \approx 220\ {\rm Hz}$ as well as a $\omega_y/(2\pi)\approx 30\ {\rm Hz}$ contribution to the trap along $\ey$.
Our experiments focused on the state space spanned by $\ket{f=1, m_F=0}$ and $\ket{f=1, m_F = 1}$ that served as our spin states $\ket{\uparrow}$ and $\ket{\downarrow}$, respectively. 
A large bias magnetic field $B_0 \approx 978\ \mu{\rm T}$ along $\ey$ provided sufficient quadratic contribution to the Zeeman shift for us to resonantly couple $\ket{\uparrow}$ and $\ket{\downarrow}$ whilst being far off resonance to $\ket{f=1, m_F=-1}$.
As seen in Fig.~\ref{fig:setup}{\bf a}, we induced spin-orbit coupling with strength $\p0 = (2\pi\hbar /\lambda_R)\times \sin(\theta)$ by Raman-coupling $\ket{\uparrow}$ and $\ket{\downarrow}$ with a pair of $\lambda_R = 790.03\ {\rm nm}$ laser beams intersecting with half-angle $\theta = 16.5$ degrees giving $\omega_0/(2\pi) \approx 300\ {\rm Hz}$, and prepared the SOC ground state (see Methods).
The spin-dependent loss in Eq.~\eqref{eq:non-Hermitian_Hamiltonian} was created by simultaneously microwave coupling $\ket{\uparrow}$ (with Rabi frequency $\Omega_\mu$) to an intermediate electronic ground state $\ket{i} = \ket{f=2, m_F = 0}$ while expelling atoms with a laser resonant between $\ket{i}$ and the $\ket{f'=3, m'_F=0}$ excited state.
This laser, aligned along $\ez$, had intensity $I$ and optical Rabi frequency $\Omega_{\rm L} =  \Gamma \sqrt{I / (2 I_{\rm sat})}$ in terms of the natural line-width $\Gamma$ and saturation intensity ${\rm I}_{\rm sat}$~\footnote{Here we use the saturation intensity suitable for the transition under consideration.}.
We chose $\Omega_\mu$ and $\Omega_{\rm L}$ to minimize the instantaneous occupation of the $f=2$ manifold, thereby mitigating density-dependent optical effects~\cite{wang2024exceptional}.

{\it Dynamical evolution} -- 
We begin by considering the dynamics of Hermitian observables.
Reference~\cite{pan2020non} showed that the response to general non-Hermitian Hamiltonians can be exotic and very different from their Hermitian counterparts.
Subject to non-Hermitian gauge field $\mathcal{B}$, the Heisenberg equation of motion (EoM) for an operator $\hat o$ becomes
\begin{align}
\label{eq:eqn_of_motion}
i\hbar \partial_t \langle \hat o \rangle &= \langle [\hat o, \hat H_{\rm h}] \rangle - \frac{i \mathcal{B}}{m^*} \langle \{\delta \hat p, \delta \hat o\} \rangle,
\end{align}
where $\delta \hat o \equiv \hat o - \langle \hat o \rangle$ and $\hat H_{\rm h}$ is the Hermitian contribution to $\hat H_{\rm nh}$ (see Methods for details). 
For example, this implies that the EoM for momentum $i\hbar \partial_t \langle \hat p \rangle = \langle [\hat p, \hat V] \rangle - 2 i \mathcal{B} \langle \delta \hat p^2 \rangle /m^*$ (i.e., the force equation), directly depends on the variance of the momentum distribution $\langle\delta \hat p^2\rangle$.
The physical origin of this ``self-acceleration'' can be understood by considering the free evolution of a Gaussian wavepacket in momentum space, with initial state shown by the black curve in Fig.~\ref{fig:setup}{\bf c}-bottom.
The anti-Hermitian contribution to the Hamiltonian, $-i\hat p {\mathcal B} / m^*$, gives a time evolution operator $\hat U_{\rm ah} = \exp[-p {\mathcal B} t / (\hbar m^*)]$ that includes loss for $p>0$ and gain for $p<0$.
As shown by the red curves (with dashed including only the imaginary vector potential, and solid adding the overall loss term $\gamma/2$), the action of this operator shifts the centroid by $-2 {\mathcal B} \Delta p^2 t/ (\hbar m^*)$ (where $\Delta p$ is the initial width of the distribution), as expected from the force equation.

In a similar manner, the EoM for the center-of-mass (CoM) position is given by
$i\hbar \partial_t \langle \hat x \rangle = i \hbar\langle \hat p \rangle/m^* - i \mathcal{B} \langle \{\delta \hat p, \delta \hat x\} \rangle /m^*$,
indicating that the CoM dynamics are influenced not only by the canonical momentum $\langle \hat p \rangle$ but also by an additional anomalous term.
This ``self displacement'' term arises from the imaginary gauge field interacting with correlated fluctuations in momentum and position.

While these EoMs effectively isolate the role of the imaginary gauge potential in dynamics, they are difficult to employ for predicting dynamics: the equations for $\langle\hat x\rangle$ and $\langle\hat p\rangle$ are not closed.
That is, they involve an infinite hierarchy of coupled equations describing ever higher order anti-commutators and operator products.
For example, computing the acceleration demands the time derivative $\partial_t \langle \{\delta \hat p, \delta \hat x\} \rangle$.
Using Eq.~\eqref{eq:eqn_of_motion} together with the kinetic and force equations, we obtain
\begin{equation}
\label{eq:total_acc}
    \hbar \partial_t^2 \langle \hat x \rangle = i\frac{\langle[ \hat V, \hat p]\rangle}{m^*} -\frac{4 \mathcal{B}}{(m^*)^2}\left\langle \delta  \hat p^2\right\rangle+i \frac{\mathcal{B}}{m^*}\langle\{\delta  \hat x,[ \hat V,  \hat p]\}\rangle + \dots\ ,
\end{equation}
which is only valid for short times (see the Methods).

\begin{figure}[tb!]
\centering
\includegraphics{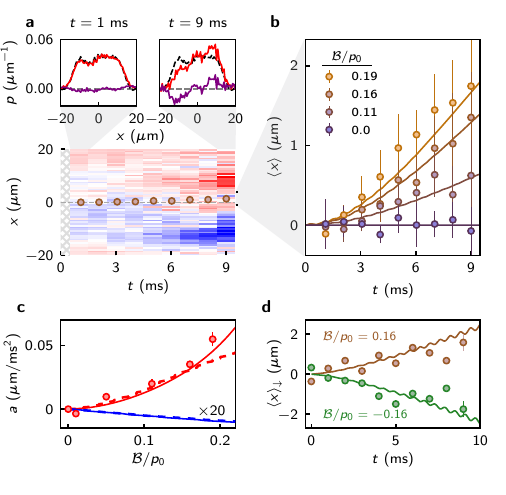}
\caption{Nonreciprocal transport.
{\bf a} Time-dependence of 1D atomic density at $\mathcal{B}/p_0=0.16$.
Top: density profiles at $t=1$ and $t=9~\mathrm{ms}$ (red) along with the $t=0$ profile (black-dashed) and their difference $\Delta n$ (purple).
Bottom: $\Delta n$ as function of evolution time along with the measured CoM position (markers).
In \textbf{b, c} and \textbf{d}, markers are experimental data and solid curves are GPE simulations.
{\bf b} Center of mass time evolution for a range of $\mathcal{B}/p_0$.
Each point reflects the average of about 10 measurements and the uncertainties are the standard error of the mean.
{\bf c} Initial acceleration as a function of $\mathcal{B}/p_0$.
Markers result from fits to the experimental data in {\bf a}; solid curves are simulated acceleration with (red) and without (blue, magnified by 20$\times$) atomic interactions.
Dashed curves plot the corresponding EoM-predicted acceleration averaged for the first 6~ms.
{\bf d} Reversed gauge field. Single spin transport $\langle x \rangle_{\downarrow}$ at $|\mathcal{B}/p_0|=0.16$.
}
\label{fig:fig3}
\end{figure}

{\it Nonreciprocal transport} -- 
After preparing the BEC in the SOC ground state, we simultaneously applied the microwave and optical fields and allowed the system to evolve for up to $9\ {\rm ms}$.
Using the high NA microscope pictured in Fig.~\ref{fig:setup}{\bf b}, we separately measured the in-situ $\ket{\uparrow}$ and $\ket{\downarrow}$ density distributions via partial transfer absorption imaging~\cite{ramanathan2012partial}, and obtained the total density from their sum (Fig.~\ref{fig:fig3}{\bf a}-top).
As time progresses, a clearly visible displacement develops in the differential density $\Delta n(t) = n(t) - n(0)$, leading to the small shift in $\langle x \rangle$ (markers in Fig.~\ref{fig:fig3}{\bf a}-bottom).
The expanded view in fig.~\ref{fig:fig3}{\bf b} shows that the CoM initially evolves proportionally to $t^2$.
We extracted the acceleration from quadratic fits to the early-time dynamics (up to $9\ {\rm ms}$), and plot it as a function of the imaginary gauge potential $\mathcal{B}/p_0$ in Fig.~\ref{fig:fig3}{\bf c} derived from observed loss rate $\gamma$ with Eq.~\eqref{eq:gauge}.
The solid blue curve, representing the acceleration predicted by the non-interacting Hamiltonians thus far discussed, bears no resemblance to our data.

Going beyond this description, our BEC also has interactions that are described by the Gross-Pitaevskii equation (GPE).
This effectively augments the Hamiltonian with a mean field energy $g |\psi(x)|^2$, which is proportional to the local atomic density $|\psi(x)|^2$ and an interaction strength $g$.
For our experimental parameters, GPE simulations of the continuum HN model [Eq.~\eqref{eq:imaginary_gauge_potential}] are in agreement with those of full two-band SOC Hamiltonian [Eq.~\eqref{eq:non-Hermitian_Hamiltonian}]; therefore, wherever possible we compare our experimental data to the suitable interacting continuum HN model.

The solid curves in Fig.~\ref{fig:fig3}{\bf b} represent GPE simulations of our experiment performed with no adjustable parameters and show excellent agreement with our observations.
Likewise, the GPE-predicted acceleration (solid red curve in Fig.~\ref{fig:fig3}{\bf c}) also closely matches our measured data.
Additionally, the dashed curves confirm that the closed-form expression for the short-time acceleration [Eq.~\eqref{eq:total_acc}] is valid for both the non-interacting and interacting cases [i.e., using $V \rightarrow V+g|\psi|^2$ in Eq.~\eqref{eq:total_acc}] up to $\mathcal{B}/p_0\approx 0.18$, beyond which higher-order terms become non-negligible (see Methods).

The stark contrast between the single-particle and GPE accelerations arise from the markedly different spatial modes of these two cases.
On one hand, the noninteracting particle in a box has a cosinusoidal ground state $\propto \cos(\pi x/L)$ that is hardly affected by the $0.9\ \mu{\rm m}$ box-trap projection resolution.
On the other hand, the BEC wavefunction remains essentially uniform until it drops rapidly to zero at the box edge over a length scale determined by an interplay of the $\xi \approx 0.3\ \mu{\rm m}$ healing length and the NA-limited sharpness of the trap's edge.
This interaction-driven change enhances both the self-acceleration and self-displacement contributions to the $\langle \hat p \rangle$ and $\langle \hat x \rangle$ EoM's.
Because both $\xi$ and the projection resolution are much smaller than $L$, the interacting system's momentum-space wavefunction is broadened, thereby enhancing the self-acceleration $\propto \langle\hat p^2\rangle$.
Furthermore, interactions extend the wavefunction into regions where the trap potential varies rapidly, amplifying the self-displacement contribution $\propto \langle\{\delta \hat x,[ \hat V, \hat p]\}\rangle$ and thus increasing the acceleration.

Thus far, we have modified only the magnitude of the imaginary gauge potential $\mathcal B$ (which is $\propto\gamma$); this is because in our loss-only system the decay rate $\gamma$ is strictly positive, so changes to it cannot affect the sign of $\mathcal B$. 

To invert the sign of $\mathcal B$, we coupled $\ket{\downarrow}$ rather than $\ket{\uparrow}$ to the lossy subspace, thereby replacing $\mathcal{P}_{\uparrow}$ with $\mathcal{P}_{\downarrow}$ in Eq.~\eqref{eq:non-Hermitian_Hamiltonian} and flipping the sign of $\mathcal B$ in Eq.~\eqref{eq:gauge}~\footnote{An equivalent sign change follows from swapping the frequencies of two Raman laser beams since $\mathcal{B} \!\propto\! p_0$.}.
Fig.~\ref{fig:fig3}{\bf d} illustrates the effect of this change: because both the self-acceleration and self-displacement are $\propto\mathcal B$, reversing its sign inverts the nonreciprocal transport.
Unlike in Fig.~\ref{fig:fig3}{\bf a}-{\bf c}, we now focus on $\langle x \rangle_\downarrow$, the CoM position of only the $\ket{\downarrow}$ spin state. 
We do so for two reasons: firstly, this component's displacement is increased relative to the average yielding a larger signal, and secondly predicting spin-resolved motion requires the full two-band SOC.
The data are consistent within its uncertainties with the result of this two-band GPE calculation (solid curves), however, the simulation exhibits high-frequency oscillations, resulting from coherent oscillations between the two SOC bands. 
These confirm that small deviations from the single-band (adiabatic) approximation exist but remain undetectable within our current experimental sensitivity.

\begin{figure}[t!]
\centering
\includegraphics{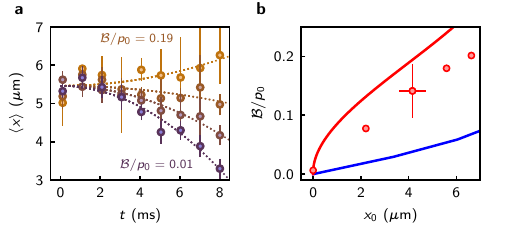}
\caption{
\textbf{Dynamical edge modes.}
{\bf a} CoM dynamics of an initially canted state for a range of ${\mathcal B}$ (markers) along with quadratic fits (dashed curves), with $\mathcal{B}/p_0$ = 0.19, 0.11, 0.05, 0.01  (top to bottom).
{\bf b} Optimal ${\mathcal B}/p_0$ from data as in {\bf a} versus CoM position $x_0$ (markers, single error bar is representative of all data).
Curves show calculations with (red) and without (blue) interactions.
}
\label{fig:edge}
\end{figure}

{\it Suppressed non-Hermitian skin effect} -- 
Both non-Hermitian terms --- $i\mathcal B$ and $i\hbar\gamma/2$ --- can be eliminated  from Eq.~\eqref{eq:imaginary_gauge_potential} by an ``anti-unitary gauge transformation'' $\hat A = \exp({\mathcal B} \hat x/\hbar - \gamma t / 2)$.
The transformed wavefunction $\ket{\psi^\prime} = \hat A \ket{\psi}$ obeys the usual Schrödinger equation, making it straightforward to compute and understand eigenstates and dynamics.
For example, the noninteracting ground state immediately leads to the non-Hermitian solution $\propto\cos(\pi x/L) \exp(-{\mathcal B} x/\hbar)$; this state, exponentially localized to the boundary, directly manifests the NHSE~
\footnote{We note, however, that this anti-unitary gauge transformation is incompatible with periodic boundary conditions.
}.

Expecting a similar outcome for our homogeneously trapped BEC, we prepared BECs with an initially ``canted'' density distribution and then abruptly switched $\gamma$ to a non-zero value.
Figure~\ref{fig:edge}{\bf a} shows that the CoM position undergoes uniform acceleration that depends systematically on ${\mathcal B}$; the dashed curves are quadratic fits that yield the acceleration. 
From the zero crossing of this acceleration, we identify the value of $\mathcal{B}$ with no net motion.
Data of this type, collected using different initial CoM positions, are shown in Fig.~\ref{fig:edge}{\bf b}. 
It might be tempting to interpret these configurations as equilibrium NHSE states; however, they deviate markedly from the single-particle solution (blue).

Indeed, applying the anti-unitary gauge transformation to the GPE shows that our system's ground state is expected to be described by a Thomas-Fermi–like uniform density distribution, leaving only small vestiges of the NHSE at the BEC's low-density periphery~\cite{GaugeFootnote}.
In line with recent theory~\cite{zhang2020skin,kawabata2022many,zheng2024exact}, our numerics show that the CoM displacement of the GPE ground state is non-zero, but greatly reduced compared to the non-interacting case, owing to the interaction energy cost of changing the ground state mode.

The red curve in Fig.~\ref{fig:edge}{\bf b} plots the results of a full time-dependent GPE simulation, showing significantly improved agreement with experiment.
These simulations show that the observed ``static'' configuration is actually a dynamical effect---somewhat analogous to macroscopic self trapping~\cite{Albiez2005} of BECs in double-well potentials---here the effects of self-acceleration and self-displacement are counterbalanced by the BECs repulsive interactions.

\begin{figure*}[bt!]
\includegraphics{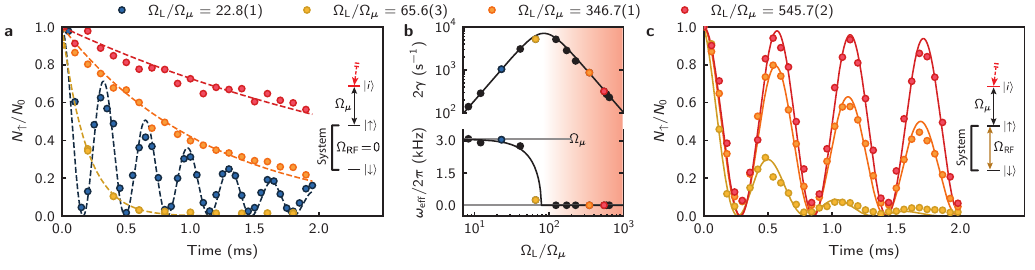}
\caption{
Transition from coherent to quantum Zeno regime.
{\bf a} Data with $\Omega_{\rm RF} = 0$ showing fractional population in $\ket{\uparrow}$ versus time for four values of $\Omega_{\rm L}$, all with $\Omega_\mu = 2\pi\times 3.08(4)\ \rm {kHz}$.
Solid curves are fits as described in the text.
{\bf b} Lifetime and oscillation frequency extracted from the fits as in {\bf a}, plotted as a function of $\Omega_{\rm L}/\Omega_{\mu}$. 
Solid curves result from time-domain master equation simulations (no free parameters), with $\omega_{\rm eff}$ and $\gamma$ obtained from fits as in {\bf a}.
{\bf c} Fractional population in $\ket{\uparrow}$ versus time, now with rf coupling $\Omega_{\rm RF} = 2\pi\times 1.20(5) \rm\ kHz$.
Solid curves show the predictions of the non-Hermitian Hamiltonian (no free parameters).
}
\label{fig:fig4}
\end{figure*}

{\it Validity of method} --
The behavior of open quantum systems---governed by the quantum master equation---can be approximated by non-Hermitian Hamiltonians only up to the first so-called quantum jump, which, for the data presented here, occurs in about twice the $26\ {\rm ns}$ lifetime of $\Rb87$'s $5{\rm}P_{3/2}$ excited state.
Nevertheless, we observe non-Hermitian evolution on millisecond timescales, raising an immediate question: how can a non-Hermitian picture remain valid for such long durations?

We experimentally explored this question in a simplified configuration using RF rather than Raman to induce $\ket{\uparrow}\!\leftrightarrow\!\ket{\downarrow}$ coupling~\footnote{For these data set the bias field to $B_0 = 670\ \mu{\rm T}$}.
Beginning with a $\spinup$ BEC, we abruptly applied combinations of optical, microwave, and RF fields for a duration $t$, and monitored the time-evolving populations $N_{\uparrow}$, $N_{\downarrow}$, and $N_{i}$.

We begin with the minimal scenario of no rf coupling (i.e., $\Omega_{\rm RF} = 0$) illustrated in Fig.~\ref{fig:fig4}{\bf a}, where the system state-space is just $\ket{\uparrow}$.
The markers show the time evolution of the normalized population $N_\uparrow / N_0$ for a range of optical Rabi frequencies $\Omega_{L}$.
When $\Omega_{\rm L}/\Omega_\mu \ll 1$, the system exhibits damped Rabi oscillations between $\ket{\uparrow}$ and $\ket{i}$, while for $\Omega_{\rm L}/\Omega_\mu \gg 1$, 
the evolution instead yields simple exponential decay.
In a standard three-level master equation description, a quantum jump (spontaneous decay) returns the atom from $\ket{e}$ to $\ket{i}$, at which time the non-Hermitian approximation fails. 
In our shallow trap, however, the recoil momentum imparted during spontaneous decay ejects the atom so that it never returns, effectively removing any jumped atoms from observation.
This mechanism resolves the question: no observed atom has undergone a quantum jump, making a non-Hermitian description possible for all times.

The resulting non-Hermitian Hamiltonian for the reduced system subspace, $\hat H_{\rm min} = -i \hbar \gamma {\mathcal P}_\uparrow$ predicts exponential decay (with rate $2\gamma$), and therefore requires $\Omega_{\rm L} / \Omega_\mu \gg 1$.
The dashed curves in Fig.~\ref{fig:fig4}{\bf a} show fits describing damped oscillatory behavior via $\exp(-2\gamma t) \cos^2(\omega_{\rm eff} t/2)$ to our data.
Fig.~\ref{fig:fig4}{\bf b} collects the results of such fits (markers) and compares them with parameters obtained from master equation simulations featuring quantum jumps only from $\ket{e}$ to $\ket{v}$ (curves, computed with no free parameters). 
In both cases, the oscillation frequency drops to zero when  $\Omega_{\rm L} = 2\Omega_{\mu}$, marking a sharp transition from damped oscillations to exponential decay.
In the exponential-decay regime (shaded), $\gamma$ decreases with increasing $\Omega_{\rm L}$, manifesting the quantum Zeno effect~\cite{itano1990quantum}.

These data fully calibrate the non-Hermitian contribution to the system Hamiltonian.
To verify this calibration, we introduce RF coupling in the Zeno regime (Fig.~\ref{fig:fig4}{\bf c}) and compare the observed dynamics with the predictions of our non-Hermitian model (solid curves). 
No free parameters are used: the decay rates come from Fig.~\ref{fig:fig4}{\bf b}, and the RF Rabi frequency was independently measured. 
Our results thus confirm that the non-Hermitian description is quantitatively accurate beyond the nominal quantum jump timescale.

{\it Discussion and outlook} --
Here we realized an imaginary gauge potential using a spin-orbit coupled BEC with laser-induced loss, and demonstrated that such a framework can be valid for all times.
This implementation serves as a specific example of a more general scenario~\cite{Zhou2025}, in which a subspace $\ket{S}$, evolving under otherwise Hamiltonian evolution, experiences quantum jumps into an uncoupled subspace $\ket{V}$.

In our system, repulsive interactions have the apparently contradictory effects of enhancing self-acceleration while simultaneously suppressing the NHSE; this naturally raises questions about the impact of these mean-field interactions on spectral topology and the non-Hermitian bulk-boundary correspondence.
While our results focused on CoM variables, the underlying GPE simulations show that this is accompanied by the formation of rapidly oscillatory shock-waves, with wavelength below our imaging resolution.
These patterns are reminiscent of those predicted in the context of non-Hermitian analog gravity simulations~\cite{Farrell2023}.

The original Hatano and Nelson initially considered an interplay between Hermitian disorder and an imaginary gauge potential~\cite{hatano1996localization}; this configuration can easily be realised in cold-atom experiments.
Moreover, by patterning the loss beam disorder can be added to the imaginary gauge potential~\cite{huang2020anderson,tzortzakakis2020non}.
Moving beyond simple single-particle or mean-field descriptions introduces core conceptual questions, because loss and gain reflect transitions between Fock spaces of total atom number, rather than simply modifying the wavefunction normalization.
In this context, defining non-Hermitian many-body topological invariants is a key challenge~\cite{kawabata2022many,zhang2020skin}.
Despite these challenges, new strongly correlated physics is predicted in non-Hermitian quantum systems.
For instance, asymmetric fermionization dynamics have been theorized in the Tonks-Girardeau regime of the 1D Bose gas~\cite{mao2023non}.
In lattice systems, combining nonreciprocal hopping with Hubbard interactions may stabilize novel phases, such as non-Hermitian Mott insulators and many-body localized states~\cite{kawabata2022many,zheng2024exact,hamazaki2019non,zhang2020skin}.

\begin{acknowledgments}
\noindent
{\it Acknowledgments} -- The authors thank R.~Ravi and K.~Weber for carefully reading the manuscript, as well as Q.~Zhou and W.~Xu for stimulating discussions.

\noindent
{\it Funding} -- This work was partially supported by the National Institute of Standards and Technology; the Quantum Leap Challenge Institute for Robust Quantum Simulation (OMA-2120757); and the Air Force Office of Scientific Research Multidisciplinary University Research Initiative ``RAPSYDY in Q'' (FA9550-22-1-0339).

\noindent
{\it Author Contributions} -- J.T. and E.D.M.G. conducted the experiment, performed theoretical work, and analyzed data; J.T., E.D.M.G., and M.Z. developed, implemented, and maintained the experimental apparatus. J.T. and E.D.M.G. conceived of the experiment with input from I.B.S.; I.B.S. provided mentorship and obtained funding. All authors substantially participated in the discussion and the writing of the manuscript.

\noindent
{\it Competing Interests} -- The authors declare that they have no
competing financial interests.


\noindent
{\it Correspondence} -- Correspondence and requests for materials
should be addressed to I.B.S.~(email: ian.spielman@nist.gov).
\end{acknowledgments}

\bibliography{main}

\begin{thebibliography}{40}
\bibitem [{Gau()}]{GaugeFootnote}%
  \BibitemOpen
  \href@noop {} {}\bibinfo {note} {In this gauge, the interaction coefficient
  becomes spatially dependent, $g(x) = g \exp(2 {\mathcal B} x/\hbar)$ leading
  to a position-dependent healing length $\xi$ that slightly shifts $\langle x
  \rangle$ towards the edge with larger $g$.}\BibitemShut {Stop}%
\end{thebibliography}

\clearpage

\appendix

\onecolumngrid

\section{Experimental parameters}

The Raman coupling parameters were chosen to maximize the observable impact of $\mathcal{B}$.
Firstly, we selected the Raman coupling strength to be $\Omega \gtrsim 4 \omega_0$: large enough for the ground SOC band to be nominally harmonic with effective mass $m^*\gtrsim m$, yet small enough to minimize spontaneous emission driven heating.
Recalling that $\omega_0 \propto p_0^2$, the remaining contributions to  $\mathcal{B}$ in Eq.~\eqref{eq:gauge} are inversely proportional to $p_0$ and therefore a decreasing function of intersection angle $\theta$.
Our final choice of $\theta = 16.5$ degrees yields an easily observable signal while still maintaining a wide range of momenta for which Eq.~\eqref{eq:gauge} is valid.

Our SOC experiments began with spin-polarized BECs in $\ket{\downarrow}$ detuned by $\delta = 157~\omega_0$ from Raman resonance.
We adiabatically transferred the BEC to the SOC ground state (i.e.,  with $p=0$ and in the lowest SOC band) by first ramping the intensity of Raman beams from zero to $\Omega = 5.7(1) \omega_0 = 2\pi\times1.68(3)\ {\rm kHz}$ in $110\ {\rm ms}$ and then ramping detuning $\delta$ to zero in $150\ {\rm ms}$.
The system was then rendered non-Hermitian by abruptly applying the microwave and optical couplings at $t=0$, giving a loss rate $\gamma \in [0, 608]\ s^{-1}$ and $|{\mathcal B}|/p_0 \in [0, 0.19]$.

\section{Experimental data analysis}

\begin{figure}[h]
\centering
\includegraphics{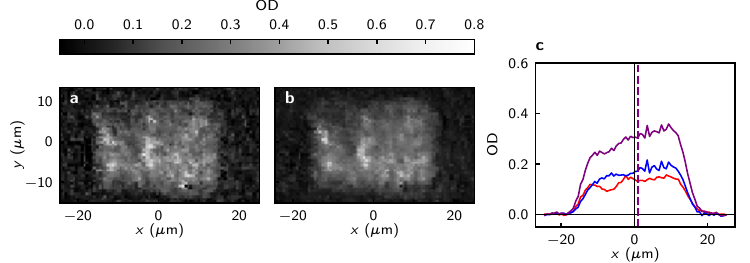}
\caption{Example data for $\mathcal{B}/p_0=0.11$ and $t=4~\mathrm{ms}$. 
{\bf a, b} Optical depth of atoms initially in $\ket{\uparrow}$ measured via partial transfer absorption imaging.
The image in {\bf b} reflects the average of 10 measurements.
{\bf c} Integrated optical depth ${\rm OD}_{\uparrow}$ (red),  ${\rm OD}_{\downarrow}$ (blue) and ${\rm OD}_{\rm tot}$ (black).
Two dashed vertical lines are $x=0$ (grey) and CoM position (purple).
}
\label{fig:fig_SM}
\end{figure}

This section outlines our analysis process for the CoM position, starting with individual images.
We measure the 2D density of each spin component by: (1) snapping off the Raman lasers; (2) then driving a $\sim \pi/4$ pulse from either $\ket{\uparrow}$ or $\ket{\downarrow}$ to $\ket{F=2, m_F=0}$; lastly (3), a resonant probe laser is used to absorption-image the transferred atoms on the $\ket{F=2}\rightarrow\ket{F'=3}$ transition for $\tau=20~\mathrm{\mu s}$.
The optical depth measured by this technique for  $\ket{\uparrow}$ is shown in Figure~\ref{fig:fig_SM}{\bf a} and {\bf b} before and after averaging, respectively.
These illustrate the flat potential along $\ex$ and the harmonically dominated potential along $\ey$.

After separately imaging both spin-components, we average along $\ey$ (orthogonal to the Raman-recoil axis) to obtain the 1D optical depths ${\rm OD}_{\uparrow,\downarrow}$, along with the total ${\rm OD}_{\rm tot}={\rm OD}_\uparrow+{\rm OD}_\downarrow$ (Fig.~\ref{fig:fig_SM}{\bf c}).
We then obtain the CoM position via $\langle x \rangle = \int dx \ {\rm OD}_{\rm tot}(x) \ x / \int dx\ {\rm OD}_{\rm tot}(x)$.

\section{Operator expectation values}

We consider the general time evolution of observables $\langle \hat o \rangle$.
For non-Hermitian evolution, the  expectation value of observables
\begin{align}
\langle \hat o \rangle &= \frac{\bra{\psi}\hat o \ket{\psi}}{\braket{\psi|\psi}}
\end{align}
includes an explicit normalization term.
This leads to a second term in the derivative
\begin{align}
\partial_t \langle \hat o \rangle &= \frac{\partial_t \bra{\psi}\hat o \ket{\psi}}{\braket{\psi|\psi}} - \langle \hat o \rangle \frac{\partial_t \braket{\psi|\psi}}{\braket{\psi|\psi}}.
\end{align}
At this point we introduce the time evolution operator $\hat U(t) = \exp(-i \hat H t / \hbar)$ as usual on the right hand side and take derivatives
\begin{align*}
i\hbar \partial_t \langle \hat o \rangle &= \frac{\bra{\psi}\left(\hat o \hat H - \hat H^\dagger \hat o\right) \ket{\psi}}{\braket{\psi|\psi}} - \langle \hat o \rangle \frac{\bra{\psi} \left(\hat H - \hat H^\dagger\right) \ket{\psi}}{\braket{\psi|\psi}}.
\end{align*}
It is convenient to break $\hat H = \hat H_{\rm h} +  \hat H_{\rm ah}$ into Hermitian $\hat H_{\rm h} = (\hat p^2 - {\mathcal B}^2)/(2m^*) + V(\hat x)$ and anti-Hermitian $ \hat H_{\rm ah} = -i {\mathcal B} \hat p /m^*$ terms to arrive at the general expression
\begin{align}
i\hbar \partial_t \langle \hat o \rangle &= \langle [\hat o, \hat H_{\rm h}] \rangle + \langle \{\hat o, \hat H_{\rm ah}\}\rangle - 2 \langle \hat o \rangle \langle \hat H_{\rm ah} \rangle\nonumber\\
&= \langle [\hat o, \hat H_{\rm h}] \rangle - \frac{i \mathcal B}{m^*} \left[\langle \{\hat o, \hat p\} \rangle - 2 \langle \hat o \rangle \langle \hat p \rangle\right]\nonumber\\
&= \langle [\hat o, \hat H_{\rm h}] \rangle - \frac{i \mathcal B}{m^*} \left[\langle \{\hat o - \langle \hat o \rangle, \hat p- \langle \hat p \rangle\} \rangle \right].\label{eq:app_main_eq}
\end{align}
This recovers Eq.~(4) in the main document.

\section{Recursive equations of motion}

Simple Hermitian Heisenberg EoM's can be closed. 
For example, for a harmonic potential the evolution of $\langle\hat x\rangle$ and $\langle\hat p\rangle$ can be described just by a pair of linear differential equations.
This is not the case for non-Hermitian Hamiltonian evolution, as we explicitly show below.

Obtaining $\langle \hat x \rangle$ as function of $t$ using the kinetic equation (i.e., setting $\hat o = \hat x$)
\begin{align}
    \label{eq:app_x_eqn}
    i\hbar \partial_t \langle \hat x \rangle = i \hbar \frac{\langle \hat p \rangle}{m^*} - i \frac{\mathcal{B}}{m^*} \langle \{\delta \hat p, \delta \hat x\} \rangle
\end{align}
requires both $\langle \hat p \rangle$ and $\langle \{\delta \hat p, \delta \hat x\} \rangle$.
We therefore use the force equation (i.e., setting $\hat o = \hat p$)
\begin{align}
    \label{eq:app_p_eqn}
    i\hbar \partial_t \langle \hat p \rangle = \langle [\hat p, \hat V] \rangle - 2 i \frac{\mathcal{B}}{m^*} \langle \delta \hat p^2 \rangle.
\end{align}
to define $\langle \hat p \rangle$, and find that these dynamics depend on $\langle \delta \hat p^2 \rangle$: yet another degree of freedom.

Equation~\eqref{eq:app_x_eqn} also depends on $\langle \{\delta \hat p, \delta \hat x\} \rangle$, which can again be obtained from Eq.~\eqref{eq:app_x_eqn},
\begin{align}
    \label{eq:app_eom3}
    i\hbar \partial_t\langle\{\delta \hat p, \delta \hat{x}\}\rangle = 2i\hbar \frac{\langle \delta \hat{p}^2\rangle}{m^*} + \langle \{\delta \hat{x},[\hat{p}, \hat{V}]\}\rangle - \frac{i \mathcal{B}}{m^*}\langle \{\delta \hat{p},\{\delta \hat{p}, \delta \hat{x}\}\} \rangle,
\end{align}
where we noted that $\langle \{\delta \hat p, \delta \hat x\} \rangle = 0$ at $t=0$ and used
\begin{align}
\left[\{\delta \hat{p}, \delta \hat{x}\}, \frac{\hat{p}^2}{2 m^*}+\hat{V}\right] = 2i\hbar \frac{\hat{p} \delta \hat{p}}{m^*}+\{[\delta \hat{p}, \hat{V}], \delta \hat{x}\}.
\end{align}

The last equation above introduced two new Hermitian contributions and one non-Hermitian contribution.
Again, the two Hermitian contributions are determined by the initial wavefunction, but we need to plug $o = \{[\delta \hat{p}, \hat{V}], \delta \hat{x}\}$ into Eq.~\eqref{eq:app_main_eq} to obtain information about the non-Hermitian anticommutator term.
For now, as the main text suggests, the first-order approximation is made by cutting off the recursion here.
We combine Eq.~\eqref{eq:app_eom3}, \eqref{eq:app_p_eqn}, \eqref{eq:app_x_eqn} and ignore
\begin{align}
    - \frac{i \mathcal{B}}{m^*}\langle \{\delta \hat{p},\{\delta \hat{p}, \delta \hat{x}\}\} \rangle
\end{align}
to derive the CoM acceleration
\begin{align}
    \hbar \partial_t^2 \langle \hat x \rangle = i\frac{\langle[ \hat V, \hat p]\rangle}{m^*} -\frac{4 \mathcal{B}}{(m^*)^2}\left\langle \delta  \hat p^2\right\rangle+i \frac{\mathcal{B}}{m^*}\langle\{\delta  \hat x,[ \hat V,  \hat p]\}\rangle
\end{align}
which was examined to agree well with the observed experimental data at short times.
In the limit $t \rightarrow 0$, simple integration obtains
\begin{align}
    \langle \hat x \rangle = i\frac{\langle[ \hat V, \hat p]\rangle}{2 \hbar m^*} t^2 + \lambda \left( -\frac{2\Delta {p}}{m^*} t + i \frac{\langle\{\delta  \hat x,[ \hat V,  \hat p]\}\rangle}{2 \hbar \Delta {p}} t \right).
\end{align}
Here, the dimensionless small expansion parameter $\lambda$ is defined
\begin{align}
    \lambda := \frac{\mathcal{B} \Delta {p}}{\hbar m^*} t
\end{align}
where $\Delta {p} = \sqrt{\langle \delta \hat{p}^2 \rangle}$.

Furthermore, when one goes beyond the first-order approximation, the recursive structure of the equations of motion is revealed.
The next order requires the introduction of the equation for the ignored term
\begin{align}
    \label{eq:app_eom4}
    i\hbar \partial_t\langle\{\delta \hat{p},\{\delta \hat{p}, \delta \hat{x}\}\}\rangle=
    4 i\hbar \frac{\langle \hat{p}(\delta \hat{p})^2\rangle}{m^*} +
    4\langle\{\delta \hat{p}, \delta \hat{x}[\hat{p}, \hat{V}]\}\rangle -
    \frac{i \mathcal{B}}{m^*}\langle\{\delta \hat{p},\{\delta \hat{p},\{\delta \hat{p}, \delta \hat{x}\}\}\}\rangle,
\end{align}
where we note
\begin{align}
    \left[\{\delta \hat{p},\{\delta \hat{p}, \delta \hat{x}\}\}, \frac{\hat{p}^2}{2 m^*}+\hat{V} \right] = 4i\hbar \frac{(\delta \hat{p})^2 \hat{p}}{m^*}+ 4\{\delta \hat{p}, \delta \hat{x}[\delta \hat{p}, \hat{V}]\}.
\end{align}
Henceforth, besides the ordinary Hermitian terms, each EoM introduces a new anti-commutator.
Eventually, we need to generate an infinite hierarchy of equations to calculate these higher-order anti-commutators $\{\delta \hat{p},\{\delta \hat{p}, ...\{\delta \hat{p}, \delta \hat{x}\}\}...\}$, which are essential to complete the system.

In a higher-order approximation, we cut off the hierarchy at order $n$ defined by $\{\underbrace{\delta \hat{p},\{\delta \hat{p}, ...\{\delta \hat{p}}_n, \delta \hat{x}\}\}\}$, which provides an approximate solution to the dynamics with our finite experiment time $t$ and small imaginary gauge potential $\mathcal{B}$.
In addition, these EoMs have an even-odd structure.
For example, Eq.~\eqref{eq:app_x_eqn} evaluates zero at $t=0$, and yet Eq.~\eqref{eq:app_p_eqn} is finite.
Likewise, Eq.~\eqref{eq:app_eom4} is zero and Eq.~\eqref{eq:app_eom3} is finite.
In other words, the early-time dynamics of $\langle \hat{x} \rangle$ is a constant acceleration, $\langle \{\delta \hat{p}, \delta \hat{x}\} \rangle$ is linear in $t$, and $\langle \{\delta \hat{p},\{\delta \hat{p}, \delta \hat{x}\}\} \rangle$ is quadratic in $t$.
More generally, these equations indicate that the anti-commutator $\{\underbrace{\delta \hat{p},\{\delta \hat{p}, ...\{\delta \hat{p}}_n, \delta \hat{x}\}\}\}$ is a linear function of $t$ when $n$ is odd and a quadratic function when $n$ is even.
This observation suggests we can only cut off the hierarchy where $n$ is odd.

Returning to our first-order approximation, a calculation to the next order is essential to validate it.
The statement above concludes Eq.~\eqref{eq:app_eom4} alone is insufficient because it is zero at $t=0$.
Hence one needs to calculate
\begin{align}
    \label{eq:app_eom5}
    i\hbar \partial_t\langle \hat{p}(\delta \hat{p})^2\rangle = \langle [\hat{p}(\delta \hat{p})^2,\hat{V}]\rangle - \frac{i \mathcal{B}}{m^*} \langle 2 \hat{p}(\delta \hat{p})^3\rangle
\end{align}
\begin{align}
\begin{split}
    i\hbar \partial_t\langle\{\delta \hat{p}, \delta \hat{x}[\hat{p}, \hat{V}]\}\rangle = &\langle -\frac{\hat{p}^4 \hat{V} \delta \hat{x}}{2 m^*} + \frac{i \hat{p}^3 \hat{V} \hbar }{2 m^*} - \frac{\hat{p}^3 \hat{V} \delta \hat{x} \delta \hat{p}}{m^*} + \frac{\hat{p}^3 \hat{V}\hat{p} \delta \hat{x}}{m^*} + \frac{i \hat{p}^2 \hat{V}\hat{p} \hbar }{2 m^*} + \frac{\hat{p}^2 \hat{V}\hat{p} \delta \hat{x} \delta \hat{p}}{m^*} \\
    &- \frac{\delta \hat{x} \hat{p}\hat{V}\hat{p}^3}{m^*} + \frac{\delta \hat{x} \delta \hat{p} \hat{p}\hat{V}\hat{p}^2}{m^*} - \frac{5 i \hat{p}\hat{V}\hat{p}^2 \hbar }{2 m^*} + \frac{2 i \hat{p}\hat{V}\hat{p} \delta \hat{p} \hbar }{m^*} + \frac{\delta \hat{x} \hat{V}\hat{p}^4}{2 m^*} - \frac{\delta \hat{x} \delta \hat{p} \hat{V}\hat{p}^3}{m^*} \\
    &+ \frac{3 i \hat{V}\hat{p}^3 \hbar }{2 m^*} - \frac{2 i \delta \hat{p} \hat{V}\hat{p}^2 \hbar }{m^*} - i \hat{p} \hat{V}^2 \hbar + 2 \hat{p} \delta \hat{x} \delta \hat{p} \hat{V}^2 + 2 \delta \hat{x} \delta \hat{p} \hat{p}\hat{V}^2 \\
    &- 2 \hat{p} \delta \hat{x} \hat{p}\hat{V}^2 - i \hat{p}\hat{V}^2 \hbar - 2 \hat{V} \delta \hat{x} \hat{V}\hat{p}^2 + 2 i \hat{V}\hat{p}\hat{V} \hbar - 4 \hat{V}\hat{p}\hat{V} \delta \hat{x} \delta \hat{p} + 4 \hat{V}\hat{p}\hat{V}\hat{p} \delta \hat{x} \rangle\\
    & - \frac{i \mathcal{B}}{m^*}\langle\{\delta \hat{p}, \{\delta \hat{p}, \delta \hat{x}[\hat{p}, \hat{V}]\}\}\rangle.
\end{split}
\end{align}
\begin{align}
\begin{split}
    i\hbar \partial_t \langle\{\delta \hat{p},\{\delta \hat{p},\{\delta \hat{p}, \delta \hat{x}\}\}\}\rangle = & \langle\frac{4 i \hbar (\delta \hat{p})^3 \hat{p} }{m^*} + 4 \hat{p}^2 \hat{p}\hat{V} \delta \hat{x} - 8 i \hbar \hat{p}^2 \hat{V} + 12 \hat{p}^2 \hat{V} \delta \hat{x} \delta \hat{p} - 12 \hat{p}^2 \hat{V}\hat{p} \delta \hat{x} - 4 \hat{p} \delta \hat{x} \hat{V}\hat{p}^2 \\
    &- 16 i \hbar \hat{p}\hat{V} \delta \hat{p} + 12 \delta \hat{p}^2 \hat{p}\hat{V} \delta \hat{x} + 12 \delta \hat{x} \hat{p}\hat{V}\hat{p}^2 + 16 i \hbar \hat{p}\hat{V}\hat{p} - 24 \hat{p}\hat{V}\hat{p} \delta \hat{x} \delta \hat{p} + 12 \delta \hat{x} \delta \hat{p} \hat{V}\hat{p}^2 \\
    &- 8 i \hbar \hat{V}\hat{p}^2 + 16 i \hbar \hat{V}\hat{p} \delta \hat{p} - 12 \delta \hat{p}^2 \hat{V}\hat{p} \delta \hat{x} \rangle -\frac{i \mathcal{B}}{m^*}\langle\{\delta \hat{p}, \{\delta \hat{p},\{\delta \hat{p},\{\delta \hat{p}, \delta \hat{x}\}\}\}\}\rangle.
\end{split}
\end{align}
Note that we kept the third-order anti-commutator.
To simplify expressions, we make an approximation to ignore quantum fluctuations which collapses all the commutators, and let replacements $\delta p \rightarrow p, \delta \hat{x} \rightarrow \hat{x}$ for small $t\rightarrow 0$
\begin{align}
    i\hbar \partial_t\langle \hat{p}(\delta \hat{p})^2\rangle =& - \frac{i \mathcal{B}}{m^*} \langle 2 \hat{p}^4\rangle \\
    i\hbar \partial_t\langle\{\delta \hat{p}, \delta x[\hat{p}, \hat{V}]\}\rangle =&- \frac{i \mathcal{B}}{m^*}\langle\{\delta \hat{p}, \{\delta \hat{p}, \delta \hat{x}[\hat{p}, \hat{V}]\}\}\rangle\\
    =&- \frac{i \mathcal{B}}{m^*} \langle 4\hbar^2 V'' + 2\hbar^2 \hat{x} V''' + 2\hbar^2 \hat{x} V'' \partial_x \rangle\\
    i\hbar \partial_t \langle\{\delta \hat{p},\{\delta \hat{p},\{\delta \hat{p}, \delta \hat{x}\}\}\}\rangle = & \frac{4i\hbar \langle \hat{p}^4 \rangle}{m^*}.
\end{align}
We used the fact that the initial distribution is symmetric in the phase space, and hence $\langle [\hat{p}(\delta \hat{p})^2,\hat{V}]\rangle =0 , \langle\{\delta \hat{p}, \{\delta \hat{p},\{\delta \hat{p},\{\delta \hat{p}, \delta \hat{x}\}\}\}\}\rangle=0$.
Combining these, one derives
\begin{align}
    -\frac{\mathcal{B}}{m^*}\langle \{\delta \hat{p},\{\delta \hat{p}, \delta \hat{x}\}\} \rangle = \left(\frac{\mathcal{B} t}{m^*}\right)^2  \left[ 6\frac{\langle \hat{p}^4 \rangle}{m^*} + 2i\hbar  \langle\{\delta \hat{p}, \{\delta \hat{p}, \delta \hat{x}[\hat{p}, \hat{V}]\}\}\rangle \right].
\end{align}
Together, the second-order expansion in $\lambda$ is
\begin{align}
    \langle  \hat{x} \rangle = i\frac{\langle[  \hat{V},  \hat{p}]\rangle}{2 \hbar m^*} t^2 + \lambda \left( -\frac{2\Delta {p}}{m^*} + i \frac{\langle\{\delta   \hat{x},[  \hat{V},   \hat{p}]\}\rangle}{2 \hbar \Delta {p}} \right) t
    - \frac{\lambda^3}{(\Delta {p})^3} \left( 3\frac{\langle \hat{p}^4 \rangle}{m^*} + \frac{i}{\hbar} \langle\{\delta \hat{p}, \{\delta \hat{p}, \delta \hat{x}[\hat{p}, \hat{V}]\}\}\rangle \right) t.
\end{align}
This confirms our assertion above.
Using induction, if one cuts off the recursion at the $n$-th order anti-commutator $\{\underbrace{\delta \hat{p},\{\delta \hat{p}, ...\{\delta \hat{p}}_n, \delta \hat{x}\}\}\}$, the expansion for the CoM displacement $\langle \hat{x} \rangle$ will be polynomial up to $\lambda^{n}$.

\section{Origin of the imaginary gauge potential $\mathcal{B}$}

The spin-orbit coupling Hamiltonian is analytically diagonalizable.
In this section we will extend this to the non-Hermitian Hamiltonian~\eqref{eq:non-Hermitian_Hamiltonian}.
We rewrite it as
\begin{equation}
\begin{aligned}
    \hat H_{\rm SOC} = &\frac{ (\hat{p} + p_0 \sigma_z)^2 }{2m} + \frac{\hbar \Omega}{2}\sigma_x +\frac{\hbar\delta}{2} \sigma_z - i \hbar \gamma \mathcal{P}_{\uparrow} \\
    = & \frac{ (\hat{p} + p_0 \sigma_z)^2 }{2m} + \frac{\hbar \Omega}{2}\sigma_x +\frac{\hbar (\delta-i\gamma) }{2} \sigma_z - \frac{\hbar \gamma}{2} \mathcal{I}.
\end{aligned}
\end{equation}
In the following, we consider the overall loss term as a constant offset.

We diagonalize the Hamiltonian and derive the eigenenergies
\begin{equation}
\begin{aligned}
    E_\pm = &\frac{p^2}{2m} \pm \hbar \sqrt{\left(\frac{\Omega}{2}\right)^2+\left(\frac{p_0 p}{m\hbar} + \frac{\delta}{2} -i\frac{\gamma}{2}\right)^2} \\ 
    = & \frac{p^2}{2m} \pm \hbar \sqrt{\left(\frac{\Omega}{2}\right)^2+ \left(\frac{p_0 p}{m\hbar}\right)^2 - i\frac{p_0 p}{m\hbar}\gamma - \left(\frac{\gamma}{2}\right)^2 }.
\end{aligned}
\end{equation}
To connect with our experiments, we focus on $E_-$ (the lower band) with $\delta=0$.

In a perturbative treatment, by assuming $\gamma \ll \Omega, p_0^2/2m$, we expand
\begin{equation}
\begin{aligned}
    E_- &\approx E_{-0} + i \hbar \frac{1}{2} \frac{p_0 p}{m\hbar} \gamma \left[\left(\frac{\Omega}{2}\right)^2+\left(\frac{p_0 p}{m\hbar} \right)^2\right]^{-1/2}
    + \hbar \frac{\gamma^2}{8} \left(\frac{\Omega}{2}\right)^2 \left[\left(\frac{\Omega}{2}\right)^2+\left(\frac{p_0 p}{m\hbar} \right)^2\right]^{-3/2}
\end{aligned}
\end{equation}
where
\begin{equation}
    E_{-0} = \frac{p^2}{2m} - \hbar \sqrt{\left(\frac{\Omega}{2}\right)^2+\left(\frac{p_0 p}{m\hbar} \right)^2}
\end{equation}
is the lower band energy without the spin-dependent loss term $\mathcal{P}_\uparrow$.

In our experiment, we load atoms into the ``single-minimum'' phase of the Hamiltonian, where $\Omega > 2 p_0^2/m\hbar$ and allows us to expand near $p=0$,
\begin{equation}
\begin{aligned}
   \left[\left(\frac{\Omega}{2}\right)^2+\left(\frac{p_0 p}{m\hbar} \right)^2\right]^\alpha \approx & \left(\frac{\Omega}{2}\right)^{2\alpha} \left[1 + \alpha \left(\frac{2p_0 p}{m\hbar\Omega}\right)^2\right]. \\
\end{aligned}
\end{equation}
Then we collect terms in orders of $\gamma$ and $p$
\begin{equation}
\begin{aligned}
    E_- &= p^2 \left(\frac{1}{2m} - \frac{p_0^2}{m^2\hbar\Omega} - \frac{3}{2} \frac{\gamma^2}{\Omega^2} \frac{p_0^2}{m^2\hbar\Omega} \right) + i \frac{p_0 \gamma}{m \Omega} p - \frac{\hbar\Omega}{2} + \frac{\hbar\gamma^2}{4\Omega}.
\end{aligned}
\end{equation}
From this result we can see that the non-Hermitian term modifies the effective mass 
\begin{equation}
    m^* = m \left[ 1 - \frac{2 p_0^2}{m \hbar\Omega} - \frac{3 p_0^2}{m \hbar\Omega} \left(\frac{\gamma}{\Omega}\right)^2 \right]^{-1}.
\end{equation}
In addition, we can formulate the imaginary term into an imaginary gauge potential
\begin{equation}
    \mathcal{B} = - \frac{p_0 \gamma}{\Omega} \frac{m^*}{m}.
\end{equation}

\end{document}